\documentstyle[12pt,psfig,twoside]{article}
\begin{document}
\hbadness=10000
\pagenumbering{arabic}
\pagestyle{myheadings}
\markboth{J. Letessier, A. Tounsi and J. Rafelski}{Hot hadronic matter
and strange anti-baryons}
\title{\bf HOT HADRONIC MATTER AND STRANGE ANTI-BARYONS}
\author{$\ $\\
\bf Jean  Letessier \hspace{1.5cm} \bf Ahmed Tounsi\\
$\ $\\
Laboratoire de Physique Th\'eorique et Hautes Energies\thanks{\em Unit\'e
associ\'ee au CNRS UA 280}~, Paris
\thanks{Postal address: LPTHE~Universit\'e PARIS 7, Tour 24, 5\`e \'et.,
2 Place Jussieu, F-75251 CEDEX 05.}~, France\\
$\ $\\
and\\
$\ $\\
\bf Johann Rafelski \\
$\ $\\
Department of Physics, University of Arizona\thanks{Postal
address:~Department of Physics, University of Arizona,
Tucson, AZ 85721}}
\date{}   % Deleting this command produces today's date.
\maketitle
%%%%%%%%%%%%%%%%%%%%%%%%%%%%%%%%%%%%%%%%
\begin{abstract}
\noindent{We demonstrate that both quark-gluon plasma (QGP) and hadronic
gas (HG) models of the central fireball created in S $\to$ W collisions
at 200 GeV A are possible sources of the recently observed strange
(anti-) baryons. From the theoretical point of view, the HG
interpretation we attempt remains more obscure because of the high
fireball temperature required. The thermal properties of the fireball as
determined by the particle ratios, are natural for the QGP state. We show
that the total particle multiplicity emerging from the central rapidity
region allows to distinguish between the two scenarios.}\\
\centerline{Published in {\it Phys. Lett.} B 292 (1992) 417--423.}
\end{abstract}
\vspace*{0.5cm}
\vfill{\bf PAR/LPTHE/92-23}  \hfill{\bf June 1992}\\
{\bf AZPH-TH/92-21}\eject
%%%%%%%%%%%%%%%%%%%%%%%%%%%%%%%%%%%%%%%%%%%%%%%%%%%
\section{\bf  Strange quarks sources}
In collisions of relativistic heavy nuclei, there is a substantially
enhanced production of strange hadronic particles\cite{strange}. Since
the time scale in a typical nucleus-nucleus collision is very short, it
is necessary to determine kinetically the strange particle  production
\cite{MYSTR}. Theoretical models of kinetic strangeness production
suggest that abundant strangeness is either a signature of QGP or of some
other new phase, e.g., with partially restored chiral symmetry
\cite{Brown} in which the cross section for strangeness production is
considerably enhanced. Enhancement in the total strangeness abundance
tells us that we deal with a state of matter which is either dense or
relatively long-lived or in which strangeness production cross sections
are enhanced or possesses some combination of these three factors. In
order to be more specific about the nature of the dense matter,
individual strange quark and anti-quark clusters, which may be more
sensitive to the environment from which they emerge \cite{RD87}, have to
be considered.  Recent measurements of the abundance of strange
anti-baryons and baryons at 200 GeV A collisions of S-ions with a
W-target by the WA85 collaboration \cite{WA85}constrains the properties
of the central rapidity source of these particles and it has already been
demonstrated that the particle abundances are well in agreement with a
picture of explosively disintegrating QGP fireball \cite{Raf91}.
Similarly, it has been already argued that these results are compatible
with the scenario of an equilibrium HG fireball  \cite{CS92}. Given that
both approaches (HG and QGP fireballs) can account for the presented
data, our main objective here will be to identify a difference between
QGP and HG scenarios which permits to distinguish the underlying
strangeness source despite the limited information available today. We
note that on theoretical grounds the QGP interpretation is a more
palatable one in view of the domain of the thermal parameters associated
with the fireball.

A simple distinction of these two phases derives from the inherent
difference with regard to their entropy content $\cal S$ given a fixed
and conserved property, such as baryon number content  $\cal B$ which can
be  determined experimentally.  $\cal B$ is seen as being well understood
in terms of the nucleon number of the combined system of the projectile
nucleus and the target tube of nuclear matter cut out in the collision
from the much larger target nucleus. Baryon number of the fireball can
decrease only by particle radiation in the final disintegration of the
fireball, beyond which we assume that the scattering between the
different components have ceased and the relative abundances carry the
information about the property of the source.  On the other hand, once
the pre-equilibrium reactions have been terminated, and the particle
momentum distributions have reached their thermal form, entropy
production effectively has ceased, even if a phase transition occurs from
a primordial phase to the final HG state \cite{entropy}. Hence both
baryon number and entropy content of the isolated fireball remain
constant and their ratio in a theoretical description is rather model
independent. Therefore the supplementary measurement, which will permit
to define the properties of this source is the multiplicity per
participating baryon in the fireball which is directly related to
entropy. While the hadronization of the entropy rich QGP fireball is
presently not understood, we take advantage here of the fact that in any
case a  substantially enhanced particle multiplicity must result, as
compared to the HG scenario. This can e.g arise if the QGP fireball were
to evaporate emitting hadronic particles sequentially. In contrast to
fireball entropy, the thermal fraction of energy of the fireball is
decreasing in the course of fireball evolution even though the total
energy content is conserved; it can not be therefore used for the study
of the properties of the source on a model independent way.

Implicit in the physical picture developed here is the formation of a
central and hot matter fireball in the nuclear collisions at 200 GeV
A.This reaction picture does not necessarily imply ``full stopping'' of
either baryon number or energy when the smaller projectile collides at
small impact parameter with the target, actually effectively with the
tube of matter in its path in the target nucleus. All that such a picture
presumes is that the energy and flavor content of the central rapidity
region is able to scatter several times, leading to thermal particle
spectra in $m_\bot$. Support for such source of strange particles comes
from the remarkably central production of strange particles as reported
by the NA35 collaboration \cite{NA35} even for S--S collisions and more
recently by the NA36 experiment \cite{NA36} for S $\to$ Pb. All these and
other data \cite{NA38} can be combined to show that the source of
transversely produced particles has a common apparent temperature of
$T=210\pm10$ MeV \cite{Raf92}.

%%%%%%%%%%%%%%%%%%%%%%%%%%%%%%%%%%%%%%%%%%%%%%%%%%%%%%%%%%%%%%%%%%
\section{\bf  Strange particle ratios}
The statistical variables of the system are the temperature $T$ and the
chemical potentials $\mu_i$ of the different conserved quark flavors
$u,d,s$. It is convenient to denote:
\begin{eqnarray}\mu_{\rm q}=(\mu_{\rm d}+\mu_{\rm u})/2 \ ,\quad
\delta\mu=\mu_{\rm d}-\mu_{\rm u} \ , \quad
\mu_{\rm B}=3\mu_{\rm q} \ ;
\end{eqnarray}
here, $\mu_{\rm q}$ is ``quark'' chemical potential, $\mu_{\rm B}$ is the
baryo-chemical potential and $\delta\mu$ describes the (small) asymmetry
in the number of up and down quarks due to the neutron excess in heavy
ion  collisions. It is straightforward to identify that the ratio of the
number of down and up quarks in a S $\to$ W-tube collisions is 1.09, and
this number will help fix the small value of $\delta\mu$, in dependence
on the assumed structure of the source -- in any case this asymmetry
leads to small corrections in our work and though implemented, will be
not discussed at length. We further introduce the fugacities
$\lambda_i=\exp(\mu_i/T)$ which are the tools of counting the particles.
Finally we note that it is not necessary to introduce different
fugacities (and chemical potentials) for the hadronic gas phase, as the
fugacity of each HG species is simply the product of the fugacities of
the constituent quarks, viz.
$\lambda_{\rm N}=\lambda_{\rm q}^3,\  \lambda_{\rm K}=\lambda_{\rm
q}\lambda_{\bar s}$,
{ etc}. The fact that quark flavors $u,d,s$ are separately conserved on
the scale of times of the hadronic collisions implies that their number
is conserved and the production (or annihilation) can only occur in
$X\leftrightarrow q_i\bar q_i$. In consequence the chemical potentials
for particle and anti-particle flavors are opposite to each other and
this implies $(\lambda_{\bar q_i} = \lambda_{q_i}^{-1})$.

In any statistical model for the production of particles  based on a
thermal source the (relative) probability of (formation) emission of
a(composite) particle (ignoring correlation effects) is:
\begin{eqnarray}
P\propto \prod_i g_i\ \lambda_i \ \gamma_i \ e^{-E_i/T}\ .
\label{eq7}
\end{eqnarray}
For a composite particle at energy $E=\sum_i E_i$, Eq.\,\ref{eq7} becomes
simply a phase space factor times the Boltzmann exponential $e^{-E/T}$
factor. We recall that $E=m_\bot \cosh(y)$ with
$m_\bot=\sqrt{m^2+{\bf p}^2_\bot}$ (the transverse direction is with
regard to the original collision axis) and $y$ is the particle rapidity.
The other factors in Eq.\,\ref{eq7} are:\\
a) the statistical multiplicity factors $g_i$, referring to the
degeneracy of the $i\ (=u,d,s)$ component, and characterizing  also the
likelihood of finding among randomly assembled quarks, the suitable
spin-isospin of the particle;\\
b) the product of chemical fugacities $\lambda_i$ for each constituent
quark species;\\
c) allowance is made for the approach to the absolute chemical
equilibrium by strange quarks with a factor $0\le\gamma_{\rm s}\le1$, for
the strange quark flavor content of the particle.

As the method of measurement distinguishes the flavor content, we keep
explicit the product of $\lambda_i$-factors; $\gamma_{\rm s}$ will enter
when one compares particles with different number of strange quarks
and/or anti-quarks. The difference between $\gamma_{\rm s}$ and
$\lambda_{{\rm s}}$ is that $\gamma_{\rm s}$ is the same for {\it both}
$s$ and $\bar s$ {viz.} $\gamma_{\rm s}=\gamma_{\bar s}$, as both
particles are produced together and hence their total abundance is
equally distant from `full' phase space. We assume that for light flavors
the $\gamma_{\rm q}$-factor is effectively unity considering the
collision times and the strength hadronic cross sections. We note that
this counting rule allows to describe the relative abundances of strange
baryons and anti-baryons at fixed $m_\bot$. All baryons considered here
have spin $1/2$, but they include spin $3/2$ resonances which become spin
$1/2$ states through hadronic decays. This is implicitly contained in the
counting of the particles by taking the product of the quark spin
degeneracies; since in all ratios to be considered this factor is the
same, we shall not discuss it further. When considering hyperons we must
remember that there are two different charge zero states of different
isospin $\Lambda$ and $\Sigma^0$: the experimental abundances of
$\Lambda$ and ${\bar \Lambda}$ (I=0) implicitly include, respectively,
the abundance of $\Sigma^0$ and ${\overline{\Sigma^0}}$ (I=1, I$_3$=0),
arising from the decay $\Sigma^0\rightarrow\Lambda^0+\gamma(74$ MeV);
hence the true abundances must be corrected by a factor $\simeq 2$, e.g.
$\Xi/\Lambda=2\cdot\Xi/(\Lambda+\Sigma^0)$, the latter being the observed
quantity.

Comparing {\it spectra} of particles within overlapping regions of
$m_\bot$, the Boltzmann and many statistical factors cancel, and their
respective abundances are only functions of fugacities, e.g.:
\begin{eqnarray}
R_\Xi&=&{{\overline{\Xi^-}}\over {\Xi^-}} ={{\lambda_{\rm d}^{-1}
\lambda_{\rm s}^{-2}}\over {\lambda_{\rm d}\lambda_{\rm s}^2}}\ ,\quad
R_\Lambda={\overline{\Lambda}\over \Lambda} ={{\lambda_{\rm d}^{-1}
\lambda_{\rm u}^{-1}\lambda_{\rm s}^{-1}}\over {\lambda_{\rm d}
\lambda_{\rm u}\lambda_{\rm s}}} .
\label{ratio}
\end{eqnarray}
The cascade and lambda ratios can easily be related to each other, in a
way which shows explicitly the chemical potential dependence:
\begin{eqnarray}
R_\Lambda &=& R_\Xi^2 \cdot e^{2\delta\mu/T}e^{6\mu_{\rm s}/T}\ ,\quad
R_\Xi=R_\Lambda^2 \cdot e^{-\delta\mu/T}e^{6\mu_{\rm q}/T}\ .
\label{R2}
\end{eqnarray}
Always remember that Eq.\,\ref{R2} is generally valid, irrespective of
the state of the system (HG or QGP). Since the asymmetry $\delta\mu$ is
small (see below), Eq.\,\ref{R2} determines the value of the
baryo-chemical and strange potentials to considerable precision (because
of the factor 6 in the exponents). The value emerging from such an
analysis will correspond to the conditions prevailing in the source of
the strange (anti-) baryons.

In Eq.\,\ref{ratio} above the factor $\gamma_{\rm s}\ (<1)$ did not
appear because it is the same for both strange quarks and anti-quarks. It
enters where particles of differing strangeness content are compared:
\begin{equation}
\overline{R_{\rm s}}\equiv{{\overline{\Xi^-}}\over \overline{\Lambda}}
={\overline{\Lambda} \over \overline{p}} =\gamma_{\rm s}\cdot
{{\lambda_{\rm s}^{-1}}\over{\lambda_{\rm u}^{-1}}}\ ;\hspace{1cm}
R_{\rm s}\equiv{{\Xi^-}\over{\Lambda}}= {{\Lambda} \over {p}}=
\gamma_{\rm s}\cdot{{\lambda_{\rm s}}\over{\lambda_{\rm u}}}\ .
\label{RAB2}
\end{equation}
The product of both expressions in Eq.\,\ref{RAB2} determines the unknown
quantity $\gamma_{\rm s}$:
\begin{eqnarray}
\overline{R_{\rm s}} \cdot R_{\rm s} \equiv{{\overline{\Xi^-}} \over
\overline{\Lambda}} \cdot{{\Xi^-}\over{\Lambda}}
\equiv{\overline{\Lambda} \over \overline{p}} \cdot {{\Lambda} \over {p}}
=
\gamma_{\rm s}^2 \ .
\label{RAB3}
\end{eqnarray}
The factor $\gamma_{\rm s}$ accounts for much of our ignorance about the
dynamics of strangeness formation and the approach to equilibrium of the
strange quark abundance. A value $\gamma_{\rm s}\simeq 1$ is believed to
favor QGP interpretation of the data \cite{strange}.

%%%%%%%%%%%%%%%%%%%%%%%%%%%%%%%%%%%%%%%%%%%%%%%%%%%%%%%%%%%%%%%
\section{\bf  Strange baryon/anti-baryon source}
We now recall the high $m_\bot$, central $y$ strange anti-baryon
abundances emerging from high energy nuclear collisions of 200~GeV/AS
$\to$ W. The following data is from the  CERN experiment WA85
\cite{WA85}, and we have: $R_\Xi:={\overline{\Xi^-}}/ \Xi^- = 0.39\pm
0.07$ for $2.3<y<3$ and $p_\bot>1$ GeV/c. In p-W reactions in the same
$(p_\bot,y)$ region, a smaller value  $0.27\pm 0.06$ is found.
Furthermore, $R_\Lambda:={\overline{\Lambda}/\Lambda} = 0.13\pm 0.03$ for
$2.4<y<2.8$ and $p_\bot>1$ GeV/c. Here corrections were applied to
eliminate hyperons  from cascading decays. The ratio $R_\Lambda$ for S-W
collisions is slightly smaller than for p-W collisions in the same
kinematic range. WA85 also determined
$\overline{\Xi^-}/(\overline{\Lambda}+\overline{\Sigma^0}) =0.6\pm 0.2$
and  $\Xi^-/(\Lambda+\Sigma^0)=0.20\pm0.04$ at fixed $m_\bot>1.72$ GeV.
The fact that the more massive and more strange anti-cascade exceeds at
fixed $m_\bot$ the abundance of the anti-lambda is most striking, even
considering the experimental error. These results are inexplicable
presently in terms of particle cascades \cite{Csernai}. The relative
yield of $\overline{\Xi^-}$ appears to be 5 times greater than seen in
the $p-p$ ISR experiment \cite{ISR} at higher energy.

These results imply in view of Eq.\,\ref{RAB3}:
$\gamma_{\rm s}=\sqrt{0.48\pm0.17}=0.7\pm 0.1$ -- the source is near to
absolute chemical equilibrium of strangeness. Note that the error which
comprises only the experimental measurement error comes out amazingly
small for $\gamma_{\rm s}$, considering how large the error on the
individual quantities has been. The values of the chemical potentials are
determined by the two forms of the Eq.\,\ref{R2}. Neglecting the isospin
asymmetry factor, we find:
$\mu_{\rm q}/T=0.167\cdot\ln(R_\Xi/R_\Lambda^2)=0.52\pm 0.1$, and
$\mu_{\rm s}/T=0.167\cdot\ln(R_\Lambda/R^2_\Xi)=-0.03\pm 0.06$. $\mu_{\rm
s}$ vanishes within the precision of the measurements, which indicates
exact symmetry between the produced strange and anti-strange quarks,
which can only occur in either:
\begin{enumerate}
\item a baryon number free fireball of arbitrary composition -- this is
clearly excluded by the presence of a finite $\mu_{\rm q}$;\item in a QGP
phase before hadronization;
\item in a HG at the magic point in $(T,\mu_{\rm B})$ parameter space for
which the size of the phase space for strange and anti-strange baryons
{\it accidentally} agree.
\end{enumerate}

%%%%%%%%%%%%%%%%%%%%%%%%%%%%%%%%%%%%%%%%%%%%%%%%%%%%%%

\section{\bf  Interpretation}
The QGP interpretation of the conditions determined above is natural: In
QGP we always have $\mu_{\rm s}=0$ for a fireball with zero strangeness.
This implies that we implicitly assume that the loss of strangeness due
to pre-equilibrium emission is small and/or symmetric between $s$ and
$\bar s$-quarks. Furthermore, with $\mu_{\rm d}/T= 0.54$, $\mu_{\rm
u}/T=0.51$ (allowing for the asymmetry between the $u$ and $d$ quarks) we
find for the value of $\alpha_{\rm s}=0.6$ and $T=210$ MeV that the
energy density is 2.2 GeV/fm$^3$, baryon density  0.27/fm$^3$, above
normal nuclear density and the strange quark density is 0.45/fm$^3$
(allowing for the factor $\gamma_{\rm s}=0.7$ as determined). The entropy
per baryon is ${\cal S}^{\rm QGP}/{\cal B}=46.9$ and it rises slowly to
47.1 as temperature is increased by 20 MeV. Naturally, strangeness
density scales nearly with $T^3$, while the energy density scales nearly
with $T^4$. A non-negligible portion of the energy is contained in the
strange quark pair density. A further small component is in the latent
energy of the vacuum which we took here to be $170$ MeV$^4$. We note that
while presence of $\alpha_{\rm s}$ reduces the light quark and gluon
effective degrees of freedom according to the known first order
perturbative behavior, we have left the contributions of strange quarks
to entropy, energy, etc... unaffected by the perturbative QCD
interaction.

When the value of $\alpha_{\rm s}$ is changed by $\pm0.2$ there is a
small change in the entropy per baryon: for $\alpha_{\rm s}=0.4$ we find
the value ${\cal S}^{\rm QGP}=50.5$, for $\alpha_{\rm s}=0.8$ we find
${\cal S}^{\rm QGP}=41.4$. As noted before, in principle entropy must
increase in the evolution of the QGP fireball, and in practice this
increase is small once thermal conditions are reached. The entropy
content per baryon is ultimately contained in the particle multiplicity
that emerges in the central rapidity region, mostly in form of pions. As
each baryon in a non-relativistic nucleon gas has the entropy content
\begin{equation}
{{\cal S}_{\rm B}\over {\cal B}}= 2.5+{m_{\rm B}-\mu_{\rm B} \over
T}\simeq 5.5
\end{equation}
the remaining entropy, ${\cal S}_\pi={\cal S}^{\rm QGP}-{\cal S}_{\rm B}$
is divided between the pions: in a relativistic pion gas at $T\simeq
1.5m_\pi$ we have entropy content per particle $\simeq 4$. Consequently
we find that the thermal parameters of the QGP fireball would lead to a
hadronic particle multiplicity per participating baryon which must be
above $(46.9-5.5) / 4\simeq 10.5 $ with the uncertainty in the value of
$\alpha_{\rm s}$ introducing an uncertainty of one unit in this value. As
expected, the QGP picture thus leads to a relatively high final state
multiplicity.

We now turn to discuss the possibility that the single data point we have
for $T,\mu_{\rm s},\mu_{\rm q}$ is associated with a hadronic gas
fireball. It is easy to write in the Boltzmann approximation the
partition function for the strange particle fraction of the hadronic gas,
${\cal Z}_{\rm s}$, as it has been presented in Ref.\,\cite{Raf87}.
However, given the WA85 values of the thermal parameters we are obliged
to sum over many more strange hadronic particles than was done in Eq.\,4
of Ref.\,\cite{Raf87}. We have in detail:
\begin{eqnarray}
\ln{\cal Z}_{\rm s}= { {V_{\rm h}T^3} \over {2\pi^2}}
[&&\hspace{-0.6cm}
(\lambda_{\rm s} \lambda_{\rm q}^{-1} + \lambda_{\rm s}^{-1} \lambda_{\rm
q})\gamma_{\rm s} F_K+(\lambda_{\rm s} \lambda_{\rm q}^{2} + \lambda_{\rm
s}^{-1} \lambda_{\rm q}^2)\gamma_{\rm s} F_Y\nonumber \\
&&\hspace{-0.6cm}
+(\lambda_{\rm s}^2 \lambda_{\rm q} +\lambda_{\rm s}^{-2} \lambda_{\rm
q}^{-1})\gamma_{\rm s}^2 F_\Xi+(\lambda_{\rm s}^{3} + \lambda_{\rm
s}^{-3})\gamma_{\rm s}^3F_\Omega]
\label{4a}
\end{eqnarray}
where the kaon ($K$), hyperon ($Y$), cascade ($\Xi$) and omega ($\Omega$)
degrees of freedom in the hadronic gas are included successively. The
phase space factors $F$ of the strange particles are:
\begin{eqnarray}
F_K&=&\sum_j g_{K_j} W(m_{K_j}/T);\ K_j=K,K^\ast,K_2^\ast,\ldots m\le
1650\ {\rm MeV}\ ,\nonumber\\
F_Y&=&\sum_j g_{Y_j} W(m_{Y_j}/T);\ Y_j=\Lambda,
\Sigma,\Sigma(1385),\ldots m\le 1750\ {\rm MeV}\ ,\nonumber\\
F_\Xi&=&\sum_j g_{\Xi_j} W(m_{\Xi_j}/T);\ \Xi_j=\Xi,\Xi(1530),\ldots
m\le1820\ {\rm MeV}\ ,\nonumber\\
F_\Omega&=&\sum_jg_{\Omega_j}W(m_{\Omega_j}/T);\
\Omega_j=\Omega,\Omega(2250)\ .
\end{eqnarray}
where $W(x)=x^2K_2(x)$, and $K_2$ is the modified Bessel function.

We now consider a HG fireball in which the number of $s$ and $\bar s$
quarks is equal. The condition that the total strangeness vanishes takes
the form
\begin{equation}
0=\langle s \rangle - \langle \bar s \rangle=\lambda_{\rm s}
{\partial\over{\partial\lambda_{\rm s}}}\ln{\cal Z}_{\rm s}\ .
\label{Eq6}
\end{equation}
This is an implicit equation relating $\lambda_{\rm s}$ with
$\lambda_{\rm q}$ for each given $T$. In actual calculations we have
distinguished the $u,d$ quarks and introduced as required $\lambda_{\rm
u},\lambda_{\rm d}$ in lieu of $\lambda_{\rm q}$, with the ratio of
$\lambda_{\rm d}/\lambda_{\rm u}$ determined by the requirement that the
ratio of down to up flavor is 1.09 (at temperature $T\simeq210$ MeV this
leads to $\delta\mu/\mu_{\rm q}\simeq 0.09$). In order to implement this
requirement we of course had to include in the partition function the
non-strange hadrons, which was done in the same way, with the mesons
included up to mass 1690 MeV, nucleons up to 1675 MeV and $\Delta$'s up
to 1900 MeV. We note that higher resonances would matter only if their
number were divergent as is the case in the Bootstrap approach of
Hagedorn \cite{HAG78} and the HG was sufficiently long lived to populate
all high mass resonances. We caution the reader that our empirical
approach suffers as soon as we ask questions which are dependent either
on the ever increasing mass spectrum of particles or on the proper volume
occupied by the particles \cite{TOUNSI91}. However, quantities such as
condition of zero strangeness, fixed entropy per baryon are independent
of the absolute normalization of the volume and of the renormalization
introduced by the diverging spectrum and hence can be considered in the
approach we take.

In the $\mu_{\rm B}-T$ plane the condition of zero strangeness combined
with the condition $\lambda_{\rm s}=1$ leads to the curve shown in
Fig.\,1 by the solid line. Dashed line is the case $\lambda_{\rm s}=0.95$
and dotted line $\lambda_{\rm s}=1.02$. The upper and lower boundary of
hatched area arise from $\mu_{\rm B}/T=3\cdot0.52\pm0.01$ and from the
constraint obtained from  the $K^-/\Lambda$ ratio.  We find that if we
are willing to accept a hadronic gas \cite{CS92} at temperature of
$T\simeq 200 - 210$ MeV, it could indeed be the source of strange
particles -- another puzzle in such an interpretation is the condition of
$\lambda_{\rm s}\simeq 1$ which is natural for QGP, and does not have at
present any special founding for the HG state.

However, we find that the properties of the HG and QGP fireballs are
considerably different in particular with regard to the entropy content.
Both states are easily distinguishable in the regime of values $\mu_{\rm
B},T$ shown in Fig.\,1. We find for ${\cal S}^{\rm HG}/{\cal B}
=21.5\pm1.5$. Consequently, the pion multiplicity which can be expected
from such a HG fireball is $4\pm0.5$. This is less than half of the QGP
based expectations we found above, and clearly the difference is
considerable in terms of experimental sensitivity. Checking the
theoretical sensitivity we find that the point at which the entropy of HG
and QGP coincide {\it and} strangeness vanishes {\it and} $\lambda_{\rm
s}\simeq1$ is at $T\simeq135$ MeV, $\mu_{\rm B}\simeq950$ MeV, quite
different from the region of interest here. We note that charged particle
multiplicity {\it above 600} in the central region has been seen
\cite{mult} in heavy ion collisions corresponding  possibly to a total
particle multiplicity of about 1,000, as required in the QGP scenario for
the central fireball we described above.

We thus conclude that the different models for the source of the strange
particles in S $\to$ W 200 GeV collisions can be differentiated with help
of the entropy content which corresponds to the final particle
multiplicity. Particle multiplicity per baryon of about 10 indicates pure
QGP fireball, around 4 suggests pure HG, and an intermediate value may
be taken to be indicative of a mixed and/or pre-critical phase. We recall
that the number of baryons in a zero impact parameter collision is about
110, and that in the above discussion multiplicity of strange particles
is included -- strangeness being a non-negligible component in the
particle flow. While we can not distinguish alone in terms of strange
particle high $m_\bot$ spectra the QGP and HG hypothesis for the case
$T=200-220$ MeV, we have shown assuming that the fireball has net
strangeness zero, that in the extreme conditions considered here both
hypothesis are falsifiable when studying the associated particle
multiplicities. \\

{\it ACKNOWLEDGEMENT}\/: J. R. would like to thank U. Heinz for
stimulating discussion of these results, and his co-authors and members
of Coll\`ege de France -- WA85 collaboration for their kind hospitality
in Paris.

%%%%%%%%%%%%%%%%%%%%%%%%%%%%%%%%%%%%%%%%%%%%%%%%%%%%%%
%\vfil\eject

%%%%%%%%%%%%%%%%%%%%%%%%%%%%%%%%%%%%%%%%%%%%%%%%%%%%%%%%%

\vfil\eject
\vspace*{-2cm}
\centerline{\psfig{figure=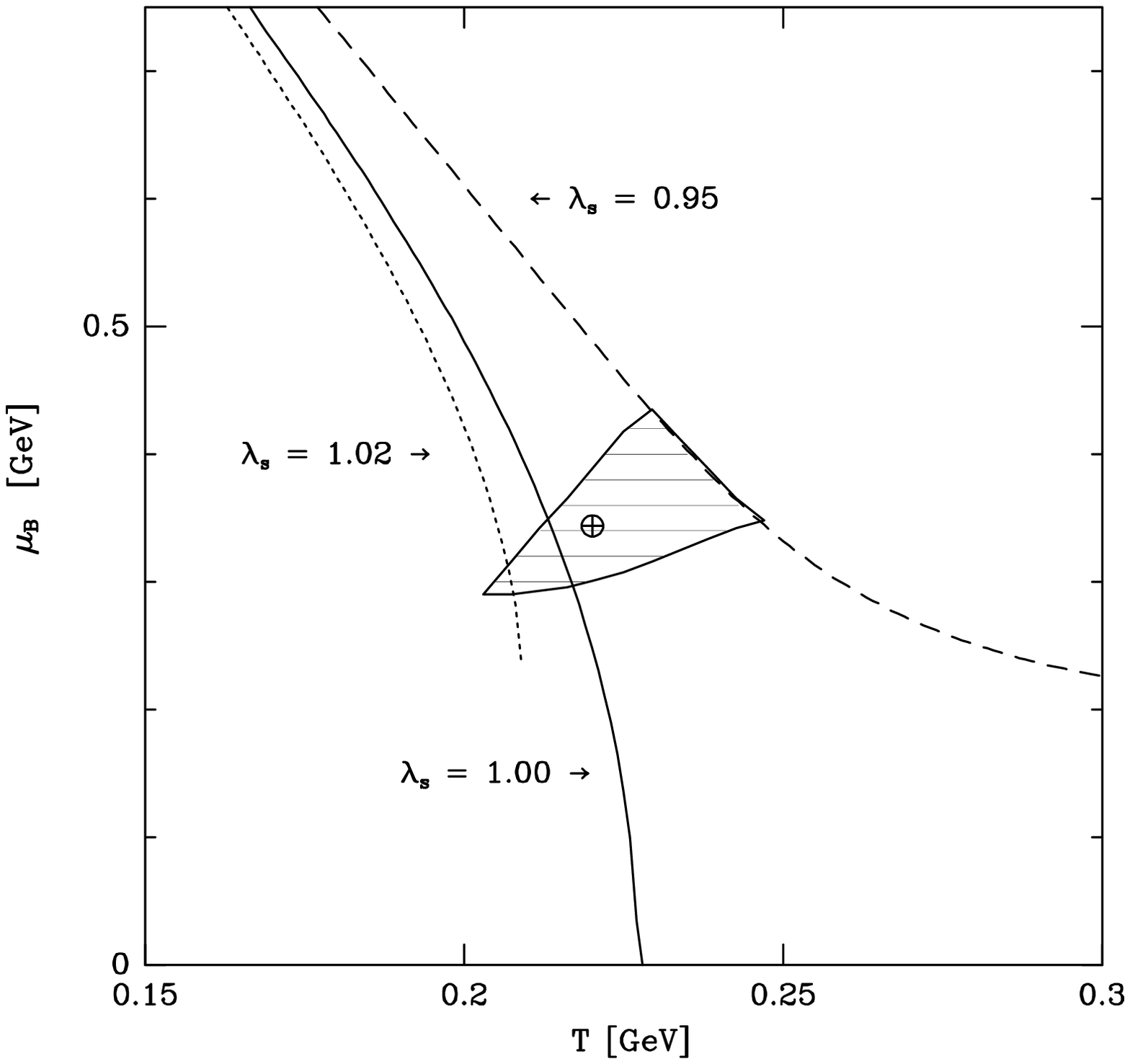,height=21.5cm}}\vspace{-2.3cm}
%\epsffile{FIG1.ps}
%\vspace*{-2cm}
\noindent{\bf Figure 1.}The solid line shows in the $\mu_{\rm B}$--$T$ plane
the condition of zero strangeness in HG fireball assuming the QGP-like
condition $\lambda_{\rm s}=1$; dashed line $\lambda_{\rm s}= 0.95$;
dotted line $\lambda_{\rm s}= 1.02$. Hatched is the region compatible
with the experimentalWA85 data. The $\oplus$ corresponds to $T =220$ MeV,
$\mu_{\rm B}=340$~MeV,the central point for QGP fireball.
\vfil\eject
\end{document}